\documentclass[11pt,english]{article}
\usepackage[T1]{fontenc}
\usepackage[latin9]{inputenc}
\usepackage{geometry}
\usepackage{babel}
\usepackage{amsthm}
\usepackage{amsmath}
\usepackage{amssymb}
\usepackage{graphicx}
\usepackage[unicode=true]
 {hyperref}

\makeatletter
\theoremstyle{plain}
\newtheorem{thm}{\protect\theoremname}
  \theoremstyle{plain}
  \newtheorem{lem}[thm]{\protect\lemmaname}
  \theoremstyle{remark}
  \newtheorem*{rem*}{\protect\remarkname}
  \theoremstyle{plain}
  \newtheorem{cor}[thm]{\protect\corollaryname}

\@ifundefined{date}{}{\date{}}
\usepackage{latexsym,color}

\@ifundefined{showcaptionsetup}{}{%
 \PassOptionsToPackage{caption=false}{subfig}}
\usepackage{subfig}
\makeatother

  \providecommand{\corollaryname}{Corollary}
  \providecommand{\lemmaname}{Lemma}
  \providecommand{\remarkname}{Remark}
\providecommand{\theoremname}{Theorem}

\begin{document}

\title{Finding the Largest Disk Containing a Query Point in Logarithmic
Time with Linear Storage%
\thanks{Work on this paper has been supported by Grant 2012/229 from the U.S.-Israeli
Binational Science Foundation, by Grant 892/13 from the Israel Science
Foundation, by the Israeli Centers of Research Excellence (I-CORE)
program (Center No.\textasciitilde{}4/11), and by the Hermann Minkowski--MINERVA
Center for Geometry at Tel Aviv University.%
}}

\author{Tal Kaminker%
\thanks{Blavatnik School of Computer Science, Tel Aviv University, Tel Aviv
69978, Israel; \protect\href{mailto:tkaminker@gmail.com}{tkaminker@gmail.com}%
} ~~~Micha Sharir %
\thanks{Blavatnik School of Computer Science, Tel Aviv University, Tel Aviv
69978, Israel; \protect\href{mailto:michas@tau.ac.il}{michas@tau.ac.il}%
}}

\date{October 11, 2013}
\maketitle
\begin{abstract}
Let ${\cal D}$ be a set of $n$ disks in the plane. We present a
data structure of size $O(n)$ that can compute, for any query point
$q$, the largest disk in ${\cal D}$ that contains $q$, in $O(\log n)$
time. The structure can be constructed in $O(n\log^{3}n)$ time. The
optimal storage and query time of the structure improve several recent
results on this and related problems, \cite{Augustine et al first ver,Augustine et al,KS}.
\end{abstract}

\section{Introduction}

Let ${\cal D}$ be a set of $n$ disks in the plane. We present a
data structure of size $O(n)$ that can compute, for any query point
$q\in\mathbb{R}^{2}$, the largest disk in ${\cal D}$ that contains
$q$, in $O(\log n)$ time. The structure can be constructed in $O(n\log^{3}n)$
time.

For simplicity, we assume general position of the disks in ${\cal D}$,
meaning, in particular, that all disks are of different sizes, and
all the $y$-coordinates of the disk centers are distinct; we also
assume distinctness of the coordinates in the directions $\pm\pi/6$.
Finally, we assume that no query point lies on the boundary of any
disk in ${\cal D}$. Degenerate situations where these assumptions
do not hold can be handled by a variety of standard techniques, such
as symbolic perturbations; see, e.g., \cite{symbolic per..}.

\paragraph{Background.}

The problem of constructing an efficient data structure for finding
the largest disk containing a query point appears to have been first
considered by Augustine et al. \cite{Augustine et al first ver} (see
also the later version of their paper \cite{Augustine et al}), as
a subproblem that arose in their solution of the problem of finding
the largest disk containing a query point, under the condition that
the disk does not contain any point of an $n$-element input point
set $P$ (the largest $P$-empty disk containing $q$). They presented
two solutions for this problem. The first solution uses a divide-and-conquer
approach that produces a data structure of size $O(n\log n)$ that
can answer a query in $O(\log^{2}n)$ time. Another solution uses
a simple sweeping technique that produces a data structure of size
$O(n^{2})$ that can answer a query in $O(\log n)$ time. A subsequent
paper by Kaplan and Sharir \cite{KS}, considering the same problem
studied in \cite{Augustine et al first ver,Augustine et al}, presents
an improved solution that uses only near-linear storage, with $O(\log^{2}n)$
query time. It is argued in \cite{KS} that the center of the largest
$P$-empty disk containing a query point $q$ must lie either on an
edge of the Voronoi diagram of $P$, or at a Voronoi vertex (assuming
that the query point lies inside the convex hull of $P$). Surprisingly,
finding the largest $P$-empty disk containing $q$ whose center lies
in the relative interior of some Voronoi edge can be done in $O(\log n)$
time using a data structure of linear size \cite{KS}. In contrast,
finding the largest $P$-empty disk containing $q$ and centered at
a Voronoi vertex (these are the Delaunay disks of $P$), with a structure
of near-linear size, could only be done in \cite{KS} in $O(\log^{2}n)$
time (using the same idea as in the first solution of Augustine et
al. \cite{Augustine et al first ver}, which also requires $\Theta(n\log n)$
storage).

This leads to a special case of the problem studied in this paper:
Given the $O(n)$ Delaunay disks of $P$, preprocess them into a data
structure of linear size, so that the largest disk containing a query
point can be found in $O(\log n)$ time. Kaplan and Sharir present
some partial results, where the query time improves to $O(k\log n),$
for any prespecified paramenter $k$, but the storage grows to $O(n^{1+\frac{1}{k}})$.

To recap, the failure of the previous works to solve the problem with
optimal query time and storage makes it an interesting challenge:
Given $n$ disks in the plane, construct a data structure of linear
size that can find, in $O(\log n)$ time, the largest disk containing
a query point.

Our algorithm meets this challenge, as specified. The only (slightly)
suboptimal performance parameter of our algorithm is the preprocessing,
which takes $O(n\log^{3}n)$ time. In particular, this improves the
result of \cite{KS}, yielding an overall algorithm that preprocesses
a set $P$ of $n$ points in the plane, in $O(n\log^{3}n)$ time,
into a data structure of linear size, that can find, in $O(\log n)$
time, the largest $P$-empty disk containing a query point. 

We note that this problem is a special case of a more general range
searching setting where ranges have priorities (in our case the size
of the disk is its priority) and we want a data structure that can
find the range of highest priority, containing a query point.

\section{The algorithm and data structure}

\subsection{Overview}

We first describe the data structure and its high-level construction.
Then we present and analyze the querying procedure. Finally, we provide
a detailed description of an efficient implementation of the construction
of the structure.

\paragraph{Construction of the data structure.}

\begin{figure}
\begin{centering}
\includegraphics[scale=0.4]{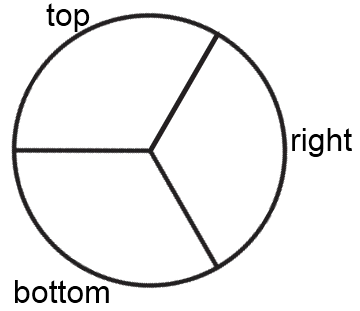}
\par\end{centering}

\caption{The partition of a disk into three equal parts; the orientation of
the three dividing radii are the same for all disks.}
\label{circle divided into 3 equal parts}
\end{figure}

The algorithm divides each disk into three equal parts, called the
\emph{right}, \emph{top}, and \emph{bottom} parts, by the radii
at orientations $\frac{\pi}{3},\,\pi\,,\textrm{ and }-\frac{\pi}{3}$
(see Figure \ref{circle divided into 3 equal parts}). The algorithm
will be run separately on all the right parts of the disks, on all
the top parts, and on all the bottom parts. The resulting substructures
will then be combined into one global structure. For simplicity, and
with no loss of generality, we will only describe the algorithm for
the right portions of the disks.

The algorithm constructs a planar map $M_{r}$, over the right portions,
with the following properties:

\paragraph*{The main properties of $M_{r}$.}
\begin{enumerate}
\item Each disk contributes at most one connnected arc to $M_{r}$.
\item No two arcs in $M_{r}$ cross each other%
\footnote{In general, an arc may terminate at a point that lies in the relative
interior of another arc.%
}.
\item For each $q\in\mathbf{\mathbb{\mathbb{R}}^{2}}$, let $D_{max}(q)$
denote the largest disk in ${\cal D}$ that contains $q$. Suppose
that $q$ lies in the right portion of $D_{max}(q)$. Then the first
arc of $M_{r}$ hit by the rightward-directed ray emanating from $q$
is the arc that $D_{max}(q)$ contributes to $M_{r}$.
\end{enumerate}
Properties (1) and (2) also hold for the two other maps. Property
(3) holds too, except that the relevant ray emanates from $q$ in
direction $2\pi/3$ (for the map of the top parts) or in direction
$-2\pi/3$ (for the map of the bottom parts).

\begin{figure}[h]
\begin{centering}
\includegraphics[scale=0.5]{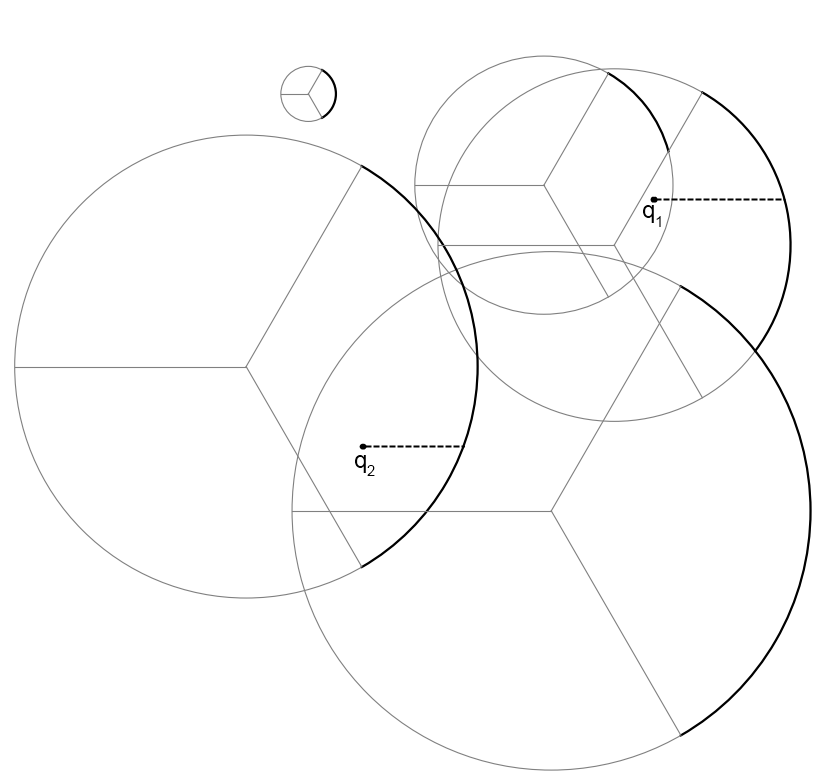}
\par\end{centering}

\caption{A set of disks and the corresponding map $M_{r}$ (consisting of the
thicker arcs). Note that the map returns the correct answer for $q_{1}$
but not for $q_{2}$, because $q_{2}$ does not lie in the right portion
of the largest disk containing it.}

\label{5-6 cirlce example}
\end{figure}

Properties (1) and (2) imply that all the maps have linear complexity.
Property (3) suggests that one should construct a point location structure
on $M_{r}$ for locating the first arc that lies straight to the right
from a query point, and construct similar structures for the other
two maps, with respect to the corresponding ray orientations $\pm2\pi/3$
mentioned above. By combining the results from all three point location
answers, one can easily find $D_{max}(q)$: it is the largest disk
that contains $q$ among the three retrieved disks. See Figure \ref{5-6 cirlce example}
for an illustration.

\subsection{Construction of $M_{r}$: High level description \label{sub:Description-of-the}}

This section presents a high-level description of the construction
of the map $M_{r}$. An actual efficient implementation of this construction
will be described later in Section \ref{sub:Constructing-the-map}.

As stated above, we divide each disk into three parts. Each part contributes
at most one arc (a subarc of its boundary arc) to the corresponding
map. As already stated, we focus only on the right part of each disk,
and on the corresponding map $M_{r}$. For each disk $d\in{\cal D}$
we denote by $T_{d}$ the right part of $d$, and by $A_{d}$ the
right arc (the arc bounding $T_{d}$). For each arc $A_{d}$, we go
over all disks larger than $d$, and each such disk $d'$ might shrink
$A_{d}$ into a potentialy smaller subarc $A_{d}^{d'}$ (possibly
eliminating it altogether). This yields, for each arc $A_{d}$, a
set of subarcs $\left\{ A_{d}^{d'}\mid\text{\ensuremath{d'}\,\ is larger than \ensuremath{d}}\right\} $,
and its intersection $A_{d}^{*}={\textstyle \underset{d'}{\bigcap}A_{d}^{d'}}$
(over the disks $d'$ larger than $d$), which is clearly a single
(possibly empty) connected subarc, is added to $M_{r}$. (This description
can trivially be turned into an $O(n^{2})$ algorithm for constructing
$M_{r}$; a more efficient near-linear construction is described in
Section \ref{sub:Constructing-the-map}.)

There are two rules for creating $A_{d}^{d'}$ from $A_{d}$ and $d'$,
which depend on the number of connected components of $A_{d}\backslash T_{d'}$.

\begin{figure}[h]
\hspace{1in}\subfloat[$A_{d}\backslash T_{d'}$ consists of one connected component.]{\includegraphics[scale=0.3]{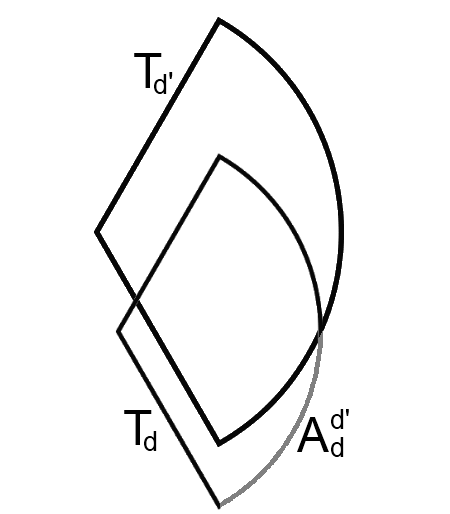}

\label{rules_sub_a}}\hspace{1in}\subfloat[$A_{d}\backslash T_{d'}$ consists of two connected components.]{\includegraphics[scale=0.3]{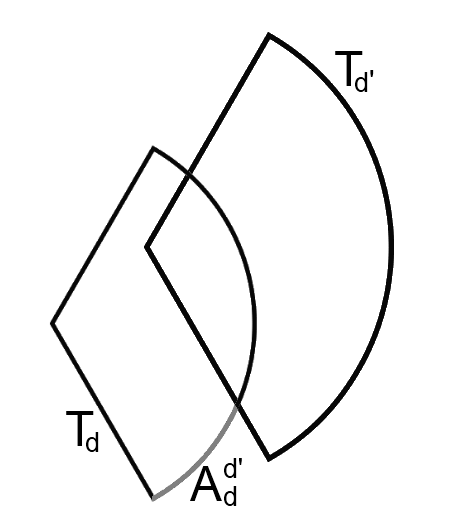}

\label{rules_sub_b}}

\caption{The two cases of the rule for constructing $A_{d}^{d'}$. In both
cases $A_{d}^{d'}$ is the grey arc (the lowest portion of $A_{d}$).}

\label{rules_figures}
\end{figure}

\paragraph{Rules for constructing $A_{d}^{d'}$.\label{sub:Rules-for-constructing}}
\begin{enumerate}
\item If $A_{d}\backslash T_{d'}$ consists of a single connected component%
\footnote{The trivial cases where $A_{d}\cap T_{d'}=\emptyset$ or $A_{d}\subset T_{d'}$
are also considered under this rule.%
} then $A_{d}^{d'}=A_{d}\backslash T_{d'}$; see Figure \ref{rules_figures}(a).
\item If $A_{d}\backslash T_{d'}$ consists of two connected components%
\footnote{It is easy to check that $A_{d}\setminus T_{d'}$ cannot have more
than two connected components, because $T_{d}$ and $T_{d'}$ are
homothetic --- see later in the paper.%
} then, if the center of disk $d$ is higher (in the $y$-direction)
than the center of $d'$, then $A_{d}^{d'}$ is the top part of $A_{d}\backslash T_{d'}$;
otherwise, $A_{d}^{d'}$ is the bottom part of $A_{d}\backslash T_{d'}$;
see Figure \ref{rules_figures}(b).
\end{enumerate}

Each of these rules creates a single (potentialy smaller) subarc of
$A_{d}$, and thus, as already noted, $A_{d}^{*}={\textstyle {\textstyle \bigcap_{d'}A_{d}^{d'}}}$
is a connected subarc of $A_{d}$. In some cases $A_{d}^{*}$ might
be empty, and then the arc $A_{d}$ does not contribute anything to
$M_{r}$, or $A_{d}^{*}$ might be equal to $A_{d}$, and then the
arc is not affected by other disks (for instance, this is always the
case for the largest disk in ${\cal D}$). This implies the first
property postulated for $M_{r}$. It is easy to see that, after the
execution of these rules on an arc $A_{d}$ and a disk $d'$ which
is larger than $d$, the resulting subarc $A_{d}^{d'}$ does not cross
the arc $A_{d'}$, although it might have one or both endpoints lying
on $A_{d'}$. Since $A_{d}^{*}\subseteq A_{d}^{d'}$, this implies
the second propery of $M_{r}$. To complete the analysis, we next
establish property (3), the main property of the map.
\begin{lem}
For each $q\in\mathbf{\mathbb{\mathbb{R}}^{2}}$, let $d=D_{max}(q)$
be the largest disk of ${\cal D}$ that contains $q$. If $q$ belongs
to the right portion $T_{d}$ of $d$, then the first arc of $M_{r}$
to the right of $q$ is $A_{d}^{*}$. \label{lem:main lemma part 1}
\end{lem}
\noindent \emph{Proof.} First we note that it suffices to prove the
lemma for the simpler case where there are only two disks, that is,
${\cal D}=\{d,d'\}$. Indeed, if the lemma were false, the ray emanating
from $q$ would not have hit $A_{d}^{*}$ first. That is, it would
have either hit first another arc $A_{d'}^{*}$, or reach infinity
without hitting any arc, and then the point where the ray hits $A_{d}$
(which always exists) must have been removed by another disk $d''$.
It is easily seen that in either case, the same situation would arise
in the presence of just $d$ and the other disk $d'$ or $d''$. 

Suppose first that $d'$ is smaller than $d$. In this case, by construction,
$A_{d}$ is not affected by $d'$, and $A_{d'}^{d}$ does not enter
the region $T_{d}$, thus rendering the lemma trivial. Suppose then
that $d'$ is larger than $d$, so $q\notin d'$. Note that in this
case we have $A_{d}^{*}=A_{d}^{d'}$. We may assume, without loss
of generality, that the center of $d$ is lower (in the $y$-direction)
than the center of $d'$; the case when the center of $d$ is higher
is handled in a fully symmetric manner. Let $r_{d'}$ and $r_{d}$
denote the respective radii of $d'$ and $d$.

We need to show that either $A_{d}^{d'}$ is the only arc that is
to the right of $q$, or else it comes before (that is, to the left
of) $A_{d'}$ (clearly, since $d'$ is larger, $A_{d'}$ appears in
its entirety in the map of only the disks $d,d'$).

First, for any point $(x_{0},y_{0})\in\mathbf{\mathbb{\mathbb{R}}^{2}}$
and any compact geometric object $A,$ define the (rightward) distance
in the $x$-direction from $(x_{0},y_{0})$ to $A$ by 
\[
dist_{x}\left((x_{0},y_{0}),\, A\right)=\min\left\{ x-x_{0}\mid(x,y_{0})\in A,\, x>x_{0}\right\} .
\]

\begin{figure}
\subfloat[The (shaded) region $L_{d'}$ of all points $p$ satisfying $dist_{x}(p,\, T_{d'})\leq r_{d'}$.]{\begin{centering}
\includegraphics[scale=0.7]{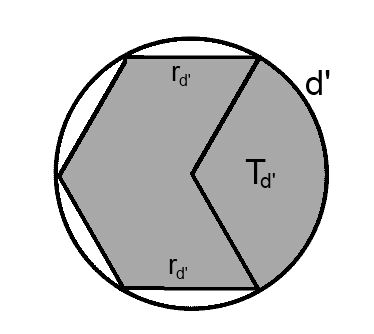}
\par\end{centering}

}\hfill{}\subfloat[The (shaded) region $K_{d'}$ of all the points $p$ satisfying $dist_{x}(p,\, A_{d'})\leq r_{d'}$.]{\begin{centering}
\includegraphics[scale=0.7]{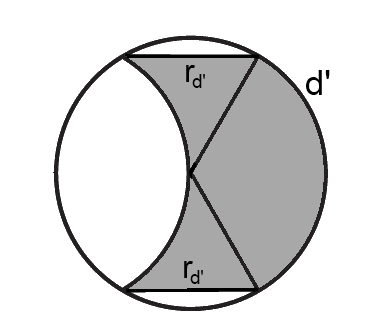}
\par\end{centering}

}

\caption{}

\label{L_d(a) and K_d(b)}
\end{figure}

The analysis relies on the following simple but crucial property.
Consider the region $L_{d'}$ of all points $p$ in the plane such
that $dist_{x}(p,\, T_{d'})\leq r_{d'}$ (see Figure \ref{L_d(a) and K_d(b)}(a)).
Then, as follows by simple geometry, $L_{d'}\subset d'$. We will
also consider the region $K_{d'}$ of all points $p$ in the plane
such that $dist_{x}(p,\, A_{d'})\leq r_{d'}$ (see Figure \ref{L_d(a) and K_d(b)}(b)).
Since $A_{d'}\subseteq T_{d'}$ we have $K_{d'}\subseteq L_{d'}\subset d'$.

Consider the rightward-directed ray $v$ emanating from $q$. Clearly
$v$ hits $A_{d}$ (since $q\in T_{d}$). Suppose that $v$ hits $A_{d'}$
before it hits $A_{d}$. This implies that $dist_{x}(q,\, A_{d'})<dist_{x}(q,\, A_{d})\leq r_{d}<r_{d'}$.
Thus $q\in K_{d'}\subset d'$, contradicting our assumption. That
is, either $v$ misses $A_{d'}$ altogether, or $v$ hits $A_{d'}$
after it hits $A_{d}.$

It remains to show that the point $I$ of intersection of $v$ with
$A_{d}$ belongs to $A_{d}^{*}=A_{d}^{d'}$. We proceed according
to the rule by which $A_{d}^{d'}$ was constructed.

\subsubsection*{Case 1: \textmd{$A_{d}\backslash T_{d'}$ consists of a single connected
component.}}

Recall that in this case $A_{d}^{d'}=A_{d}\backslash T_{d'}$. Assume
to the contrary that $I\notin A_{d}^{d'}$, so $I\in T_{d'}$. But
then $dist_{x}(q,\, T_{d'})\leq|qI|\leq r_{d}<r_{d'}$. Thus $q\in L_{d'}\subset d'$,
contrary to our assumption. Hence $I\in A_{d}^{d'}$ in this case.

\subsubsection*{Case 2: \textmd{$A_{d}\backslash T_{d'}$ consists of two connected
components. }}

\begin{figure}[h]
\subfloat[$A_{d}$ crosses the bottom edge of $T_{d'}$ twice.]{\includegraphics[scale=0.3]{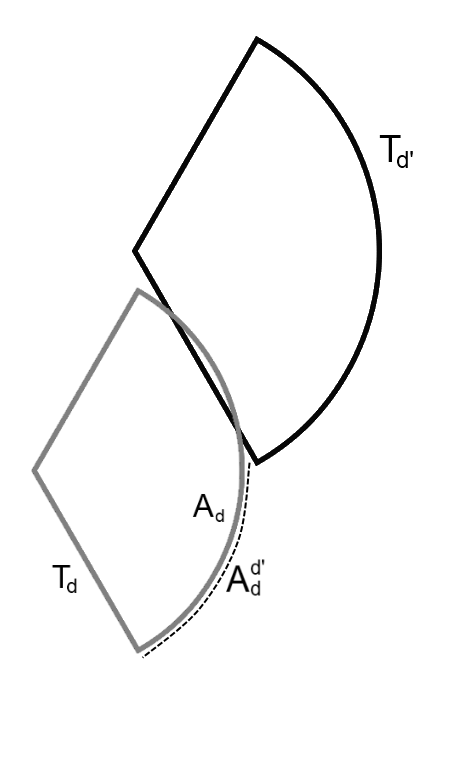}

}\hfill{}\subfloat[$A_{d}$ crosses $A_{d'}$ and the top edge of $T_{d'}$.]{\includegraphics[scale=0.3]{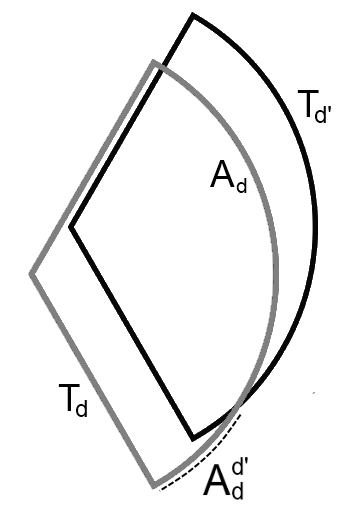}

}\hfill{}\subfloat[$A_{d}$ crosses both the top and bottom edges of $T_{d'}$.]{\includegraphics[scale=0.3]{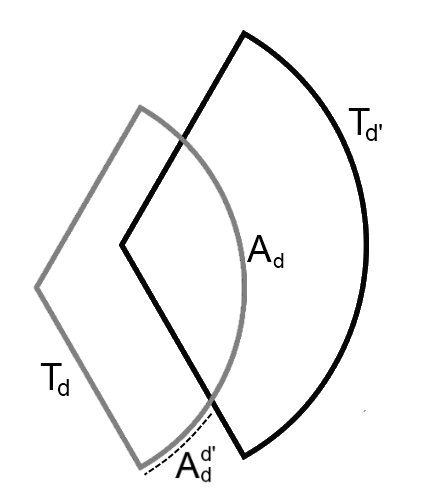}

\label{classic case 2 situation}}\hfill{}\subfloat[$A_{d}$ crosses $A_{d'}$ and the bottom edge of $T_{d'}$.]{\includegraphics[scale=0.3]{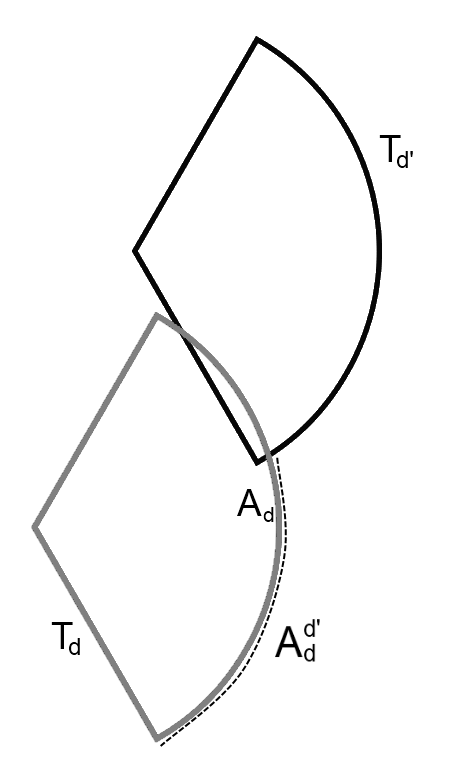}

}

\caption{The four situations of case 2; the grey disk is the smaller disk $d$.}

\label{four situations for case 2}
\end{figure}

Let $c$, $c'$ be the top and bottom vertices of $T_{d}$, respectively.
Since $T_{d}$ and $T_{d'}$ are homothetic copies of each other,
their boundaries intersect in two points (see \cite{KLPS}), both
of which lie on $A_{d}$ in this case%
\footnote{Our general position assumption allows us to assume that $\partial T_{d}$
and $\partial T_{d'}$ do not overlap (in some straight segment).%
}. This implies that either both $c$ and $c'$ lie inside $T_{d'}$
or both lie outside $T_{d'}$. If $c$ and $c'$ lie inside $T_{d'}$
then, as is easily checked, both intersection points of $\partial T_{d}$
and $\partial T_{d'}$ also lie on $A_{d'}$, and the portion $A_{d}\setminus T_{d'}$
would have to be the middle portion of $A_{d}$ (between the two intersection
points), so $A_{d}\setminus T_{d'}$ would be connected, contrary
to assumption. Using the facts that both $c$ and $c'$ lie outside
$T_{d'}$ and no more intersection points between $\partial T_{d}$
and $\partial T_{d'}$ exist except the two on $A_{d}$, we conclude
that the only part of $\partial T_{d}$ that is inside $T_{d'}$ is
the middle portion of $A_{d}$. See Figure \ref{four situations for case 2}.

\begin{figure}[h]
\begin{centering}
\includegraphics[scale=0.7]{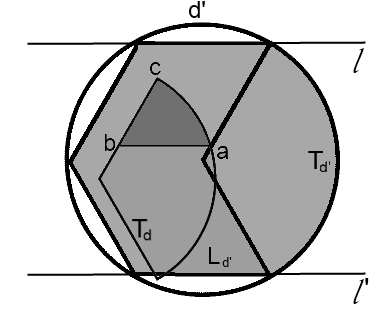}
\par\end{centering}

\caption{The triangle-like region $abc$ bounded by $ab$, $bc$, and $\overline{ac}$.
The lightly-shaded region is $L_{d'}$, as in Figure \ref{L_d(a) and K_d(b)}(a).}

\label{triangle-like diagram}
\end{figure}

Let $l$, $l'$ denote the horizontal lines touching the top and bottom
vertices of $T_{d'}$, respectively (see Figure \ref{triangle-like diagram}).
Clearly the top vertex $c$ of $T_{d}$ lies above $l'$, and since
the center of $d$ is lower than the center of $d'$, and $r_{d}<r_{d'}$,
the point $c$ has to lie below the line $l$. In conclusion $c$
lies between $l$ and $l'$. Observe also that $c$ is behind (in
the $x$-direction) $T_{d'}$, for otherwise, using the fact that
both the top and the bottom straight edges of $T_{d}$ are outside
$T_{d'}$, it would be impossible for $\partial T_{d}$ and $\partial T_{d'}$
to cross each other, contrary to assumption. Denote by $a$ the top
intersection point of $A_{d}$ and $\partial T_{d'}$. Using the fact
that $c$ is outside and behind $T_{d'}$ and between $l$ and $l'$,
and the fact that the top subarc $\overline{ac}$ of $A_{d}$ does
not cross $\partial T_{d'}$ (except touching it an $a$), we conclude
that $a$ lies on one of the straight edges of $T_{d'}$.

We next claim that the point $a$ lies above the center of $d$. Indeed,
suppose to the contrary that both intersection points of $A_{d}$
with $\partial T_{d'}$ ($a$ and the second, lower point $a'$) are
on the bottom half of $A_{d}$ (i.e., on the subarc of $A_{d}$ starting
at the bottom vertex $c'$ of $A_{d}$ up until the middle of $A_{d}$).
Since the center of $d$ is lower than the center of $d'$, both intersection
points are on the bottom half of $\partial T_{d'}$ (i.e., each lying
either on the bottom edge of $T_{d'}$ or on the bottom half of $A_{d'}$).
Since the bottom half of $A_{d}$ is the graph of a monotone increasing
function and the bottom edge of $T_{d'}$ is the graph of a monotone
decreasing function, the bottom half of $A_{d}$ and the bottom edge
of $T_{d'}$ cross each other at most once. This means that at least
one intersection point lies on $A_{d'}$, which, as we have just argued,
must be the lower point $a'$ (because $a$ lies on an edge of $T_{d'}$).
This however is impossible, because $a'$ lies to the left of $a$,
and thus it lies to the left of $A_{d'}$, contrary to assumption.
This contrudiction establishes our claim.

To proceed, we consider the possible ways in which $A_{d}$ can intersect
$\partial T_{d'}$ twice (see Figure \ref{four situations for case 2}).
As proven before, the top intersection point of $A_{d}$ and $\partial T_{d'}$
lies on one of the edges of $T_{d'}$. It is impossible that both
intersection points lie on the top edge of $T_{d'}$, for then the
center of $d$ would have to be higher than the center of $d'$, as
is easily checked. This leaves us with four subcases: Either (i) both
intersection points lie on the bottom edge of $T_{d'}$ (see Figure
\ref{four situations for case 2}(a)), or (ii) one intersection point
lies on $A_{d'}$ and one on the top edge of $T_{d'}$ (see Figure
\ref{four situations for case 2}(b)), or (iii) one intersection point
lies on the bottom edge and one on the top edge of $T_{d'}$ (see
Figure \ref{four situations for case 2}(c)), or (iv) one intersection
point lies on $A_{d'}$ and one on the bottom edge of $T_{d'}$ (see
Figure \ref{four situations for case 2}(d)). 

\begin{figure}[h]
\begin{centering}
\includegraphics[scale=0.4]{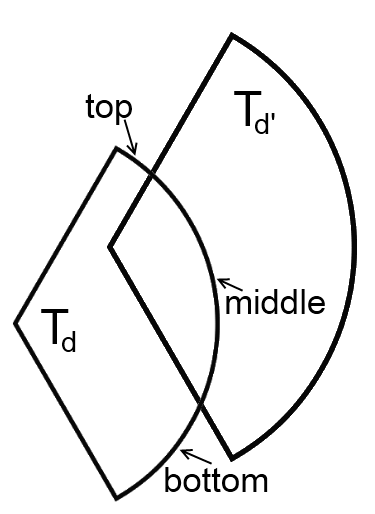}
\par\end{centering}

\caption{Decomposition of $A_{d}$ in case 2; $A_{d}^{d'}$ is the bottom subarc.}

\label{top_middle_bottom}
\end{figure}

The proof below deals with all of these situations in the same manner.
The proof-related Figures \ref{triangle-like diagram} and \ref{top_middle_bottom}
are shown for situation \ref{four situations for case 2}(c), but
the proof applies, as is, for the other cases as well. Recall that
we have assumed that the center of $d$ is below the center of $d'$
(in the $y$-direction), so that $A_{d}^{d'}$ is the bottom part
of $A_{d}$. As noted before, we only need to prove that the intersection
point $I$ of $A_{d}$ with the ray $v$ belongs to $A_{d}^{d'}$.
Suppose to the contrary that this is not the case. Then either $I$
is in the middle part of $A_{d}$ or $I$ is in the top part (see
Figure \ref{top_middle_bottom}) . An almost identical proof to the
one in case 1 shows that if $I$ is in the middle part of $A_{d}$
then $q\in d'$, contrary to our assumption. Suppose then that $I$
is in the top part of $A_{d}$.

Denote by $b$ the point of intersection of $\partial T_{d}$ and
the leftward-directed ray emanating from $a$; see Figure \ref{triangle-like diagram}.
Since $|ba|\leq r_{d}<r_{d'}$, we get that $b$ lies in $L_{d'}$.
By what has been argued above, $b$ lies on the top edge $e$ of $T_{d}$.
Since $e$ is parallel to the top left edge $e^{*}$ of $L_{d'}$,
starts (at its bottom) below $e^{*}$, and is shorter than $e^{*}$,
it follows that the top endpoint $c$ of $e$ lies in $L_{d'}$. In
conclusion, $b$ and $c$ both lie in $L_{d'}$ and outside $T_{d'}$.
The top subarc $\overline{ac}$ of $A_{d}$ is the graph of a monotone
decreasing function, both of whose endpoints lie in $L_{d'}$. It
then easily follows that the entire arc is contained in $L_{d'}$.
Since $b$ lies on the top edge of $T_{d}$, the triangle-like region
$abc$, bounded by $ab$, $bc,$ and the arc $\overline{ac}$, is
fully contained in $L_{d'}$. Since $q$ lies in this region by assumption,
it follows that $q\in L_{d'}\subset d'$, contrary to our assumption.
Hence $I$ must lie in $A_{d}^{d'}$, as claimed. 

This concludes the proof of the lemma, and thus establishes property
(3) of $M_{r}$. $\Box$

\noindent We summarize the results of this section in the following
theorem.
\begin{thm}
Given a set ${\cal D}$ of $n$ disks in the plane, one can construct
a data structure of linear size, consisting of (point location structures
for) the map $M_{r}$ and its two symmetric couterparts $M_{t}$ and
$M_{b}$, constructed respectively over the top parts and over the
bottom parts of the disks of ${\cal D}$, such that, for a given query
point $q\in\mathbb{R}^{2}$, the largest disk of ${\cal D}$ containing
$q$ is the largest of the disks that contain $q$ among the three
disks $d_{r},\, d_{t},\, d_{b}$ where $d_{r}$ is the disk whose
arc $A_{d_{r}}^{*}$ is the first arc of $M_{r}$ hit by the rightward
directed ray from $q$, and where $d_{t},\, d_{b}$ are similarly
defined for $M_{t}$, $M_{b}$, respectively, and for the respective
rays in directions $2\pi/3$, $-2\pi/3$. The query time is $O(\log n)$.
\label{thm:final of first section}\end{thm}
\begin{rem*}
The reason for dividing each disk into three parts is to ensure that
$L_{d}\subset d$ (and thus also $K_{d}\subset d$) for each disk
$d$. Both properties fail if $T_{d}$ is larger than a third of a
disk.
\end{rem*}

\subsection{Efficient Construction of $M_{r}$\label{sub:Constructing-the-map}}

As already noted, the operational definition of the map $M_{r}$,
as given in Section \ref{sub:Description-of-the}, leads to a straightforward
and simple $O(n^{2})$ algorithm for constructing the map. We now
describe a more efficient procedure for constructing $M_{r}$, which
runs in $O(n\log^{3}n)$ time. 

Fix a disk $d\in{\cal D}$, and let ${\cal D}_{d}$ denote the set
of all disks in ${\cal D}$ larger than $d$. Let ${\cal T}_{d}$
denote the collection of the right portions $T_{d'}$ of all the disks
$d'\in{\cal D}_{d}$, and let $U_{d}$ denote their union. Since the
elements of ${\cal T}_{d}$ are homothetic copies of one another,
their union has linear complexity; see \cite{KLPS,homothetic copy}.

\begin{figure}[h]
\subfloat[]{\includegraphics[scale=0.4]{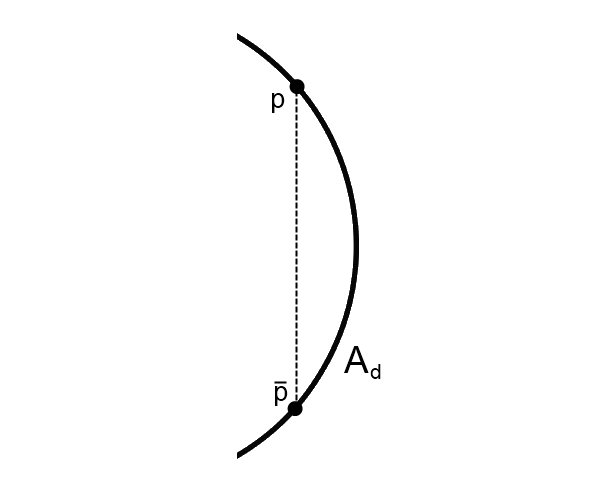}

}\hfill{}\subfloat[]{\includegraphics[scale=0.4]{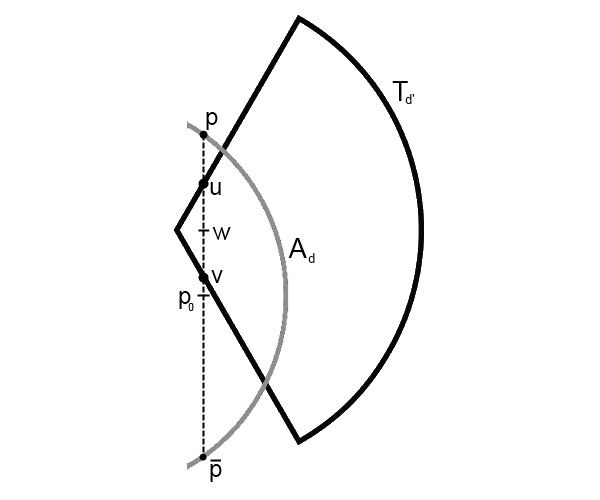}

}

\caption{(a) Point $p\in A_{d}$ and its conjugate point $\bar{p}$. (b) Illustrating
the setup in Lemma \ref{lem: negative}.}

\label{negative point}
\end{figure}

Before we proceed, we first establish the following lemma. For an
arc $A_{d}$ of some disk $d$, and a point $p\in A_{d}$ in the top
(resp., bottom) half of $A_{d}$, we define the \emph{conjugate}
point $\bar{p}$ of $p$ (with respect to $A_{d}$) to be the second
intersection point of $A_{d}$ and the vertical line through $p$;
see Figure \ref{negative point}(a).
\begin{lem}
Let $d,\, d'\in{\cal D}$ such that $d'$ is larger than $d$ and
the center of $d'$ lies above (resp., below) the center of $d$ (in
the $y$-direction). Let $p\in A_{d}\setminus T_{d'}$ be a point
in the top (resp., bottom) half of $A_{d}$. Then the conjugate point
$\bar{p}$ of $p$ (with respect to $A_{d}$) is also in $A_{d}\setminus T_{d'}$.
\label{lem: negative}
\end{lem}
\noindent \emph{Proof. }See Figure \ref{negative point}(b) for an
illustration. We may assume, without loss of generality, that the
center of $d'$ lies above the center of $d$; the complementary case
can be handled in a fully symmetric manner. In this case $p$ lies
in the top half of $A_{d}$ and $\bar{p}$ lies in the bottom half.
Suppose to the contrary that $\bar{p}$ lies in $T_{d'}$. Let $l$
be the vertical line through $p$ and $\bar{p}$. Let $u$ and $v$
denote the two intersection points of $l$ with $\partial T_{d'}$,
with $u$ lying above $v$; these points must exist, for otherwise
$\bar{p}$ would trivially lie in $A_{d}\setminus T_{d'}$. Observe
that either $u$ lies on the top edge of $T_{d'}$ and $v$ on the
bottom edge, or both points lie on $A_{d'}$. In either case, the
midpoint $w$ of $uv$ has the same $y$-coordinate as the center
of $d'$, and therefore must lie above the midpoint $p_{0}$ of $p\bar{p}$,
whose $y$-coordinate is equal to that of the center of $d$. Since
$p$ lies outside $T_{d'}$ and $\bar{p}$ lies inside, it follows
that $u$ lies between $p$ and $\bar{p}$ and $v$ lies below $\bar{p}$.
But then $w$ must lie below $p_{0}$, as is easily checked, a contradiction
that completes the proof. $\Box$

Let ${\cal D}_{d}^{(a)}$ (resp., ${\cal D}_{d}^{(b)}$) denote the
collection of all the disks $d'$ such that $d'$ is larger than $d$
and the center of $d'$ lies above (resp., below) the center of $d$
(in the $y$-direction); thus ${\cal D}_{d}={\cal D}_{d}^{(a)}\cup{\cal D}_{d}^{(b)}$.
Let ${\cal T}_{d}^{(a)}$ (resp., ${\cal T}_{d}^{(b)}$) denote the
collection of the regions $T_{d'}$, for $d'\in{\cal D}_{d}^{(a)}$
(resp., $d'\in{\cal D}_{d}^{(b)}$), and let $U_{d}^{(a)}$ (resp.,
$U_{d}^{(b)}$) denote the union of ${\cal T}_{d}^{(a)}$ (resp.,
${\cal T}_{d}^{(b)}$); in particular, $U_{d}=U_{d}^{(a)}\cup U_{d}^{(b)}$.
Lemma \ref{lem: negative} implies the following corollary.
\begin{cor}
Let $p\in A_{d}\setminus U_{d}^{(a)}$ (resp., $p\in A_{d}\setminus U_{d}^{(b)}$)
be a point in the top (resp., bottom) half of $A_{d}$. Then the conjugate
point $\bar{p}$ of $p$ (with respect to $A_{d}$) is also in $A_{d}\setminus U_{d}^{(a)}$
(resp., $A_{d}\setminus U_{d}^{(b)}$). \label{cor:negative}
\end{cor}
\noindent Set $A_{d}^{(a)}:=\bigcap_{d'\in{\cal D}_{d}^{(a)}}A_{d}^{d'}$,
and $A_{d}^{(b)}:=\bigcap_{d'\in{\cal D}_{d}^{(b)}}A_{d}^{d'}$. That
is, $A_{d}^{(a)}$ (resp., $A_{d}^{(b)}$) is the subarc of $A_{d}$
resulting from applying the rule for creating $A_{d}^{*}$ only to
the disks in ${\cal D}_{d}^{(a)}$ (resp., in ${\cal D}_{d}^{(b)}$).
By definition, $A_{d}^{*}=A_{d}^{(a)}\cap A_{d}^{(b)}$. The algorithm
uses the following lemma to construct $A_{d}^{(a)}$ and $A_{d}^{(b)}$.
\begin{lem}
$A_{d}^{(a)}$ (resp., $A_{d}^{(b)}$) is the lowest (resp., highest)
connected component of $A_{d}\setminus U_{d}^{(a)}$ (resp., $A_{d}\setminus U_{d}^{(b)}$).
\label{lem: main second section lemma}\end{lem}
\begin{rem*}
Implicitly, the lemma also asserts that $A_{d}^{(a)}$ (resp., $A_{d}^{(b)}$)
is empty if and only if $A_{d}\setminus U_{d}^{(a)}$ (resp., $A_{d}\setminus U_{d}^{(b)}$)
is empty.
\end{rem*}
\noindent \emph{Proof.} Without loss of generality, we prove the lemma
only for $A_{d}^{(a)}$, as the handling of $A_{d}^{(b)}$ is fully
symmetric. For disks $d'\in{\cal D}_{d}^{(a)}$, both rules for constructing
$A_{d}^{d'}$ boil down to the single rule that $A_{d}^{d'}$ is the
lowest connected component of $A_{d}\setminus T_{d'}$. By applying
this rule to all of the disks $d\in{\cal D}_{d}^{(a)}$, we get that
$A_{d}^{(a)}$ is disjoint from $U_{d}^{(a)}$, and since all the
subarcs $A_{d}^{d'}$, for $d'\in{\cal D}_{d}^{(a)}$, are connected,
we conclude that $A_{d}^{(a)}$ is connected and contained in a single
connected component of $A_{d}\setminus U_{d}^{(a)}$. In particular,
if $A_{d}\setminus U_{d}^{(a)}=\emptyset$ then $A_{d}^{(a)}$ is
also empty. For each $d'\in{\cal D}_{d}^{(a)}$, each of the endpoints
of the subarc $A_{d}^{d'}$ (if $A_{d}^{d'}\neq\emptyset$) is either
a vertex of $A_{d}$ or one of the intersection points of $A_{d}$
and $\partial T_{d'}$. Thus, each endpoint of the subarc $A_{d}^{(a)}$,
which is a finite intersection of such arcs $A_{d}^{d'}$, is also
either a vertex of $A_{d}$ or one of the intersection points of $A_{d}$
with some $T_{d'}$. This implies that, if not empty, $A_{d}^{(a)}$
is a maximal connected component of%
\footnote{Clearly, if $A_{d}\setminus U_{d}^{(a)}=\emptyset$ then $A_{d}^{(a)}=\emptyset$
as well; the converse implication will shortly be established.%
} $A_{d}\backslash U_{d}^{(a)}$.

Let $L$ be the lowest connected component of $A_{d}\setminus U_{d}^{(a)}$.
Clearly, if $L\neq A_{d}^{(a)}$ then either $A_{d}^{(a)}=\emptyset$
and $L\neq\emptyset$, or $A_{d}^{(a)}\neq\emptyset$ but $L$ is
a lower connected component of $A_{d}\setminus U_{d}^{(a)}$ than
$A_{d}^{(a)}$. Assume first to the contrary that $A_{d}^{(a)}\neq\emptyset$
but $L$ is a lower connected component of $A_{d}\setminus U_{d}^{(a)}$
than $A_{d}^{(a)}$. Pick points $p\in A_{d}^{(a)}$ and $p'\in L$,
so $p'$ is lower than $p$. By construction, $p\in A_{d}^{d'}$ for
every disk $d'\in{\cal D}_{d}^{(a)}$, and $p'\in A_{d}\setminus T_{d'}$.
By definition of $A_{d}^{d'}$, $p'$ must also belong to this arc.
Since this holds for all $d'\in{\cal D}_{d}^{(a)}$, $p'\in A_{d}^{(a)}$,
contrary to our assumption. Thus $L=A_{d}^{(a)}$.

Suppose then, again to the contrary, that $A_{d}^{(a)}=\emptyset$
and $L\neq\emptyset$ and let $p$ be some arbitrary point in $L$.
Notice that in order for that to happen, $p$ must be in the top connected
component of $A_{d}\setminus T_{d'}$ for some $d'\in{\cal D}_{d}^{(a)}$,
i.e., $A_{d}\setminus T_{d'}$ consists of two connected components
and $p$ is in the top one. Indeed, if this does not happen, $p$,
which lies outside all the sets $T_{d'}$, for $d\in{\cal D}_{d}^{(a)}$,
must lie in all the arcs $A_{d}^{d'}$ and thus also in $A_{d}^{(a)}$,
which is impossible. As shown in the analysis of case 2 in the proof
of Lemma \ref{lem:main lemma part 1}, the top intersection point
of $A_{d}$ and $\partial T_{d'}$ is above the center of $d$. Hence
$p$ must lie in the top half of $A_{d}$, which means that $L$ is
contained in the top half of $A_{d}$. Using Corollary \ref{cor:negative}
we conclude that the conjugate point $\bar{p}$ of $p$, which lies
in the bottom half of $A_{d}$, is in $A_{d}\setminus U_{d}^{(a)}$
as well, contradicting the fact $L$ is the lowest component of $A_{d}\setminus U_{d}^{(a)}$.

In conclusion, we have shown that $L=A_{d}^{(a)}$ in all cases. This,
and a symmetric argument for $A_{d}^{(b)}$, complete the proof of
the lemma. $\Box$ 

In other words, $A_{d}^{*}$ is the intersection of the lowest component
of $A_{d}\setminus U_{d}^{(a)}$ and the highest component of $A_{d}\setminus U_{d}^{(b)}$.

\paragraph*{Decomposability.}

\noindent We note that the observations just made are more general
in nature, and yield the following \emph{decomposability} property
of the construction of $A_{d}^{*}$. Suppose that ${\cal D}_{d}={\cal D}_{1}\cup{\cal D}_{2}\cup\cdots\cup{\cal D}_{s}$.
For each $j=1,\ldots,s$, let ${\cal D}_{j}^{(a)}$ (resp., ${\cal D}_{j}^{(b)}$)
denote the set of those disks in ${\cal D}_{j}$ whose centers lie
above (resp., below) the center of $d$, and let $A_{d;j}^{(a)}$
(resp., $A_{d;j}^{(b)}$) denote the lowest (resp., highest) connected
component of $A_{d}\setminus\bigcup{\cal D}_{j}^{(a)}$ (resp., of
$A_{d}\setminus\bigcup{\cal D}_{j}^{(b)}$). Then $A_{d}^{*}=\underset{j=1}{\overset{s}{\bigcap}}\left(A_{d;j}^{(a)}\cap A_{d;j}^{(b)}\right)$.

\paragraph*{A divide-and-conquer algorithm.}

The preceding analysis suggests the following divide-and-conquer procedure
for constructing $M_{r}$. Let ${\cal D}^{+}$ (resp., ${\cal D}^{-}$)
denote the collection of the $n/2$ larger (resp., smaller) disks
of ${\cal D}$, and let $M_{r}^{+}$ (resp., $M_{r}^{-}$) denote
the map constructed (recursively) on the right portions of the disks
in ${\cal D}^{+}$ (resp., ${\cal D}^{-}$).

Note that all the arcs in $M_{r}^{+}$ are arcs of the final map $M_{r}$
(they are not affected by the addition of the smaller disks), but
the arcs of $M_{r}^{-}$ might require some trimming to turn them
into the correct arcs in $M_{r}$, because of the effect of the larger
disks in ${\cal D}^{+}$ on them. Let $A_{d}^{-}$ be an arc in $M_{r}^{-}$,
contributed by some disk $d\in{\cal D}^{-}$. To incorporate the effect
that the disks in ${\cal D}^{+}$ have on $A_{d}^{-}$, we compute
the lowest (resp., highest) connected arc $A_{d}^{+(a)}$ (resp.,
$A_{d}^{+(b)}$) of $A_{d}\setminus\bigcup_{d'\in{\cal D}^{+}\cap{\cal D}_{d}^{(a)}}T_{d'}$
(resp., $A_{d}\setminus\bigcup_{d'\in{\cal D}^{+}\cap{\cal D}_{d}^{(b)}}T_{d'}$),
and add to the final map $M_{r}$ the arc $A_{d}^{*}:=A_{d}^{+(a)}\cap A_{d}^{+(b)}\cap A_{d}^{-}$.
The last identity follows directly from the definition: 
\[
A_{d}^{*}=\underset{d'\in{\cal D}_{d}}{\bigcap}A_{d}^{d'}=\biggl(\bigcap_{d'\in{\cal D}^{+}\cap{\cal D}_{d}^{(a)}}A_{d}^{d'}\biggr)\cap\biggl(\bigcap_{d'\in{\cal D}^{+}\cap{\cal D}_{d}^{(b)}}A_{d}^{d'}\biggl)\cap\biggr(\bigcap_{d'\in{\cal D}^{-}\cap{\cal D}_{d}}A_{d}^{d'}\biggl)=A_{d}^{+(a)}\cap A_{d}^{+(b)}\cap A_{d}^{-}\,.
\]

To complete the description of this divide-and-conquer process, we
present an efficient implementation of the construction of the arcs
$A_{d}^{+(a)}$ and $A_{d}^{+(b)}$ for the right portions of all
the disks $d\in{\cal D}^{-}$. In what follows we concentrate only
on the efficient construction of the arcs $A_{d}^{+(a)}$; computing
the corresponding arcs $A_{d}^{+(b)}$ is done in a fully symmetric
manner. The arcs $A_{d}^{*}$ are then obtained by the preceding rule,
and the construction of $M_{r}$ is completed.

Consider first the following subproblem, which arises as a major step
in the construction. We have a set ${\cal E}^{-}=\{d_{1}^{-},\ldots,d_{k}^{-}\}$
of $k$ ``small'' disks, and a set ${\cal A}^{-}=\{A_{1},\ldots,A_{k}\}$
of noncrossing arcs, so that $A_{j}$ is a subarc of the arc bounding
the right portion $T_{d_{j}^{-}}$ of $d_{j}^{-}$, for $j=1,\ldots,k$.
We also have a set ${\cal E}^{+}=\{d_{1}^{+},\ldots,d_{s}^{+}\}$
of $s$ ``large'' disks, so that all the disks of ${\cal E}^{+}$
are larger than all the disks of ${\cal E}^{-}$, and the center of
each disk of ${\cal E}^{+}$ lies above the center of every disk of
${\cal E}^{-}$ (in the $y$-direction). Let $U^{+}$ denote the union
of all the right portions $T_{d}$, for $d\in{\cal E}^{+}$. Our goal,
according to Lemma \ref{lem: main second section lemma} and the decomposability
property noted following it, is to compute, for each $j=1,\ldots,k$,
the lowest subarc $\alpha_{j}$ of $A_{j}\setminus U^{+}$.

\begin{figure}
\subfloat[]{\includegraphics[scale=0.3]{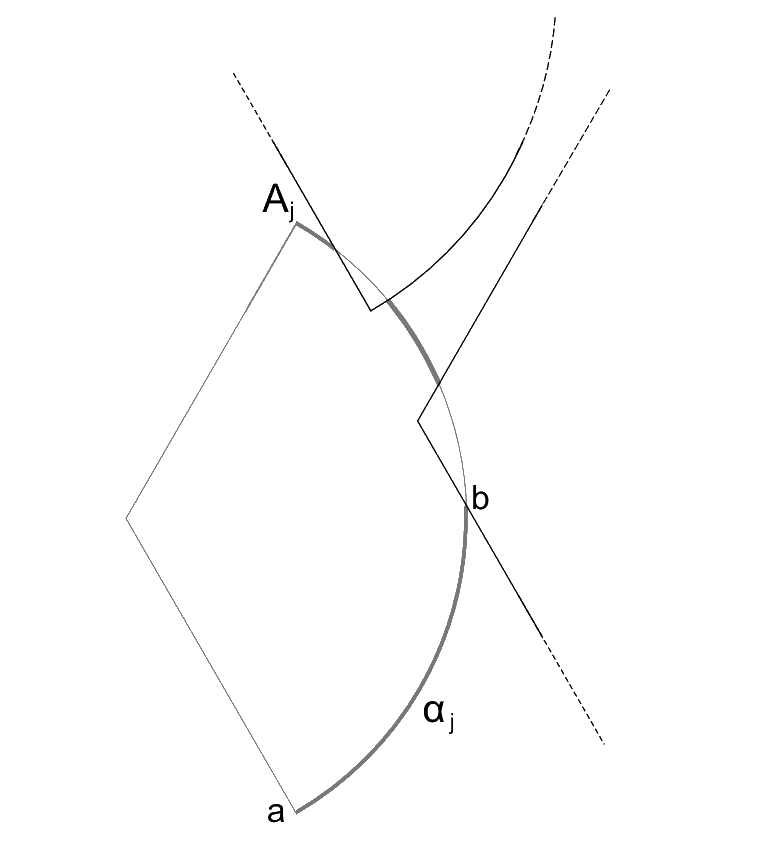}

}\hfill{}\subfloat[]{\includegraphics[scale=0.3]{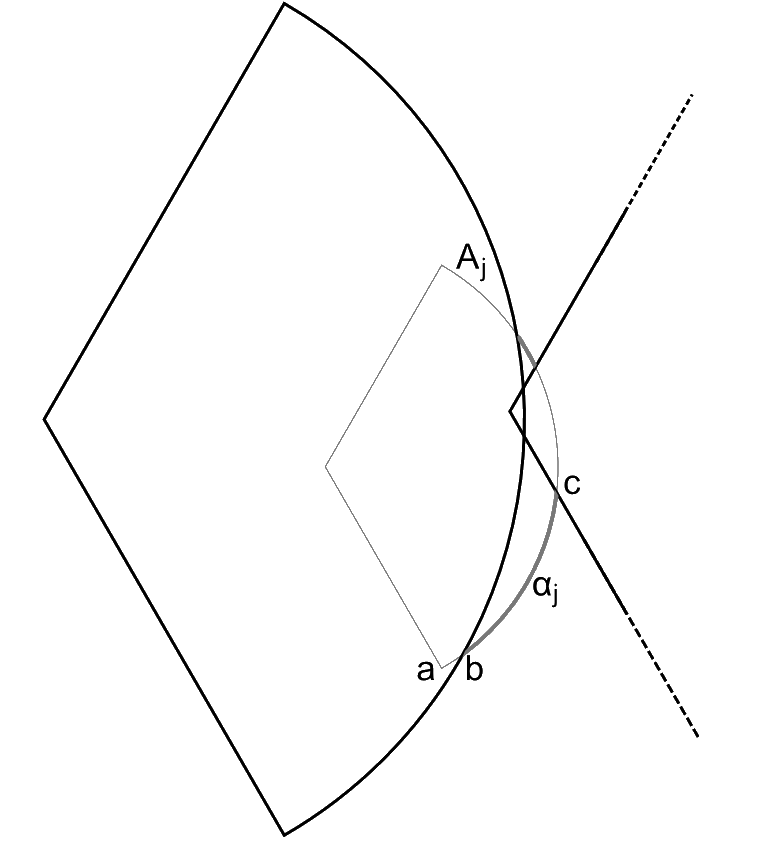}

}

\caption{(a) $\alpha_{j}$ is the arc $\overline{ab}$ when $a\notin U^{+}$.
(b) $\alpha_{j}$ is the arc $\overline{bc}$ when $a\in U^{+}$.}

\label{fig: alpha j}
\end{figure}

We first note the following operational definition of $\alpha_{j}$:
Let $a$ be the lower endpoint of $A_{j}$. If $a\notin U^{+}$ then
$a$ is also the lower endpoint of $\alpha_{j}$, and the upper endpoint
of $\alpha_{j}$ is the lowest intersection point $b$ of $A_{j}$
with $\partial U^{+}$ (or the upper endpoint of $A_{j}$ if no such
intersection exists). Otherwise, if $a\in U^{+}$, the intersection
point $b$ just defined (if it exists) is the lower endpoint of $\alpha_{j}$,
and the upper endpoint is the second lowest intersection point of
$A_{j}$ with $\partial U^{+}$ (or, again, the upper endpoint of
$A_{j}$ if no second intersection exists); if $b$ does not exist
then $\alpha_{j}=\emptyset$. See Figure \ref{fig: alpha j} for an
illustration.

We compute the union $U^{+}$, and construct on it a standard point
location data structure (see, e.g., \cite{dBCvKO}), which allows
us to determine, in logarithmic time, whether a query point lies inside
the union. As already noted several times earlier, $U^{+}$, as the
union of $s$ homothetic regions, has linear complexity (that is,
$O(s)$). The cost of constructing $U^{+}$ is discussed below, towards
the end of the presentation of the algorithm. Let ${\cal B}^{+}$
denote the set of all the $O(s)$ circular arcs and straight segments
that form the boundary of $U^{+}$.

We then run a horizontal line sweeping algorithm, in increasing $y$-direction,
on the collections ${\cal A}^{-}$ and ${\cal B}^{+}$, whose goal
is to compute, for each arc of ${\cal A}^{-}$, its lowest or two
lowest intersection points with the elements of ${\cal B}^{+}$. In
more detail, for each arc $A_{j}\in{\cal A}^{-}$, we locate the lower
endpoint $a_{j}$ of $A_{j}$ in $U^{+}$. If $a_{j}\notin U^{+}$,
it suffices to compute the first lowest intersection point, and otherwise
we need to compute the two lowest intersections.

The implementation of the sweep is straightforward; we omit here the
bulk of its description, and only give the following few comments
about its execution. First, since the arcs of ${\cal A}^{-}$ are
pairwise noncrossing, and so are the arcs and segments of ${\cal B}^{+}$,
the sweep will only encounter intersections between the elements of
${\cal A}^{-}$ and those of ${\cal B}^{+}$. For each arc $A_{j}$
of ${\cal A}^{-}$, the sweep extracts from the event priority queue
(the ``$y$-structure'') the intersection points of $A_{j}$ with
the elements of ${\cal B}^{+}$ in increasing $y$-order. As soon
as it extracts the first or second such intersection, as appropriate
for $A_{j}$, it discards $A_{j}$, removing the arc itself from the
``$x$-structure'' (the balanced search tree that represents the arcs
crossed by the sweepline), and removing any future events involving
$A_{j}$ from the queue. In this manner, the algorithm processes only
$O(k+s)$ events, at a total cost of $O((k+s)\log(k+s))$ time.

Returning to the original divide-and-conquer procedure, we would like
to apply the sweeping technique to the arcs of $M_{r}^{-}$ (playing
the role of ${\cal A}^{-}$) and to the union of the right portions
$T_{d}$ of the disks $d\in{\cal D}^{+}$ (playing the role of $U^{+}$).
However, since we do not have control over the relative positions
of the centers of the two corresponding families of disks, we need
a secondary process to control the locations of the centers.

Specifically, we sort the disks of ${\cal D}^{+}$ in increasing $y$-order
of their centers, and store them in this order at the leaves of a
balanced binary tree $Q$. For each node $v\in Q$, we consider the
subset ${\cal D}_{v}$ of the disks stored at the leaves of the subtree
rooted at $v$. Note that $\sum_{v}|{\cal D}_{v}|=O(n\log n)$. Given
a disk $d\in{\cal D}^{-}$, we can obtain, by searching in $Q$ with
the center of $d$, the subset ${\cal D}^{+}\cap{\cal D}_{d}^{(a)}$
of disks of ${\cal D}^{+}$ whose centers lie above that of $d$,
as the (disjoint) union of $O(\log n)$ subsets ${\cal D}_{v}$, and
the decomposability propery implies that it suffices to solve the
problem for each of them separately, and form the intersection of
the $O(\log n)$ resulting subarcs of $A_{d}$, to obtain the desired
subarc $A_{d}^{+(a)}$, as defined above. As a result of the searches
in $Q$ with the centers of the disks of ${\cal D}^{-}$, each node
$v\in Q$ stores a set $S_{v}$ of all the disks $d\in{\cal D}^{-}$
that use $v$ as part of their decomposition (i.e., the decomposition
of ${\cal D}^{+}\cap{\cal D}_{d}^{(a)}$). The overall size of the
sets $S_{v}$, as well as the time to construct them, is $O(n\log n)$.

Now, for each node $v\in Q$ we have a pair $(S_{v},{\cal D}_{v})$
of subsets $S_{v}\subseteq{\cal D}^{-}$ and ${\cal D}_{v}\subseteq{\cal D}^{+}$.
We recover from $M_{r}^{-}$ the arcs that correspond to the disks
in $S_{v}$, and run the sweeping algorithm described above, where
the recovered arcs play the role of ${\cal A}^{-}$, and where ${\cal D}_{v}$
plays the role of ${\cal E}^{+}$. After all the sweeps are performed,
we have computed, for each disk $d\in{\cal D}^{-}$, $O(\log n)$
arcs $A_{d}^{(1)},A_{d}^{(2)},\ldots$. As argued above, the intersection
of all these arcs is the desired $A_{d}^{+(a)}$. We then run this
procedure again, essentially reversing the direction of the $y$-axis,
to obtain the subarcs $A_{d}^{+(b)}$, for $d\in{\cal D}^{-}$, and
add $A_{d}^{*}=A_{d}^{+(a)}\cap A_{d}^{+(b)}\cap A_{d}^{-}$ to the
output map $M_{r}$. 

The total time of all the sweeps is bounded by 
\[
O\biggr(\sum_{v\in Q}(|{\cal D}_{v}|+|S_{v}|)\log(|{\cal D}_{v}|+|S_{v}|)\biggr)=O\biggr(\sum_{v\in Q}(|{\cal D}_{v}|+|S_{v}|)\log n\biggr)=O(n\log^{2}n).
\]

To complete the presentation, we next discuss the cost of constructing
the union $U_{v}^{+}=\bigcup\left\{ T_{d}\mid d\in{\cal D}_{v}\right\} $,
for all the nodes $v$ of $Q$. We construct these unions in a bottom-up
manner, from the leaves of $Q$ towards its root. The construction
of the leaves is trivial, because at each leaf the union involves
a single region $T_{d}$. Let $v$ be a non-leaf node of $Q$ with
children $w$,$z$. Then $U_{v}^{+}=U_{w}^{+}\cup U_{z}^{+}$. if
$v$ has $s$ leaves in its subtree, then each of $U_{w}^{+}$, $U_{z}^{+}$,
$U_{v}^{+}$ has $O(s)$ complexity. This means that $U_{v}^{+}$
can be constructed by a straightforward line sweeping procedure over
the overlay of $U_{w}^{+}$ and $U_{z}^{+}$, in time $O(s\log s)$.
Adding up the costs, over all nodes $v$ of $Q$, all the unions $U_{w}^{+}$
can be constructed in a total of $O(n\log^{2}n)$ time.

In conclusion, we have presented a procedure that, given two maps
$M_{r}^{-}$ and $M_{r}^{+}$, constructed respectively over the $n/2$
smaller disks of ${\cal D}$ and over the $n/2$ larger disks, merges
them into the final map $M_{r}$, in $O(n\log^{2}n)$ time. Denoting
by $T(n)$ the maximum time for the algorithm to run on a set of $n$
disks, we obtain the recurrence relation: $T(n)=2T(\frac{n}{2})+O(n\log^{2}n)$,
and thus $T(n)=O(n\log^{3}n)$.
\begin{rem*}
As described, the preprocessing algorithm uses $O(n\log n)$ space
(for maintaining all the sets $S_{v}$, ${\cal D}_{v}$). To impove
the storage requirement to linear, we construct $Q$, and the corresponding
sets $S_{v}$, ${\cal D}_{v}$, in an incremental bottom-up manner,
maintaining at each step the sets $S_{v}$, ${\cal D}_{v}$ only within
a single level of $Q$. The information concerning the sets ${\cal D}_{v}$
and the unions of their disks is easy to transfer from each level
the the next one up. To obtain the sets $S_{v}$ within a level the
simplest way is to search in $Q$ with the centers of disks of ${\cal D}^{-}$
from scratch, only until the desired level. This implementation takes
only linear space.
\end{rem*}
\noindent Putting everything together, and summarizing the statement
of Theorem \ref{thm:final of first section}, we obtain the following
summary result of this paper.
\begin{thm}
Let ${\cal D}$ be a set of $n$ disks in the plane. One can preprocess
${\cal D}$ into a data structure of linear size, in time $O(n\log^{3}n)$,
so that, for any query point $q$ we can report the largest disk of
${\cal D}$ that contains $q$ or determine that there is no such
disk, in $O(\log n)$ time.\end{thm}
\begin{rem*}
Although this is somewhat marginal, it would be interesting to reduce
the cost of the preprocessing.
\end{rem*}

\paragraph*{Acknowledgements.}

The authors wish to thank Haim Kaplan for helpful discussions and
feedback on the problem studied in this paper. We also thank indirectly
Joe Mitchell and Günter Rote, whose discussions of this problem with
Haim were a prime motivation for us to study this problem.


\begin{thebibliography}{1}
\bibitem{Augustine et al first ver} J. Augustine, S. Das, A. Maheshwari,
S. C. Nandy, S. Roy, and S. Sarvattomananda, Recognizing the largest
empty circle and axis-parallel rectangle in a desired location, in
arXiv:1004.0558, 2010.

\bibitem{Augustine et al} J. Augustine, S. Das, A. Maheshwari, S.
C. Nandy, S. Roy, and S. Sarvattomananda, Querying for the largest
empty geometric object in a desired location, in arXiv:1004.0558v2,
2010.

\bibitem{dBCvKO} M. de Berg, O. Cheong, M. van Kreveld and M. Overmars,
{\it Computational Geometry: Algorithms and Applications}, 3rd Edition,
Springer Verlag, Berlin--Heidelberg, 2008.

\bibitem{KS} H. Kaplan and M. Sharir, Finding the maximal empty disk
containing a query point, {\it Int. J. Comput. Geom. Appl.}, to appear.
Also in {\it Proc. 28-th Annu. ACM Sympos. Comput. Geom.}, 2012,
pp. 287--292

\bibitem{KLPS} K. Kedem, R. Livne, J. Pach, and M. Sharir, On the
union of Jordan regions and collision-free translational morion amdist
polygonal obstacles, {\it Discrete Comput. Geom.} 1 (1986), 59--71.

\bibitem{homothetic copy} P. K. Agarwal, J. Pach and M. Sharir, State
of the union (of geometric objects), in {\it Proc. Joint Summer Research Conf. on Discrete and Computational Geometry: 20 Years Later},
Contemp. Math. 452, AMS, 2008, pp. 9--48.

\bibitem{symbolic per..} C. K. Yap. Geometric consistency theorem
for a symbolic perturbation scheme, {\it Journal of Computer and System Sciences},
1990, 40(1), pp. 2--18.\end{thebibliography}
\end{document}